\documentclass[twocolumn,showpacs,preprintnumbers,amsmath,amssymb,prb,aps]{revtex4-1}

\usepackage{subfigure}
\usepackage{comment}
\usepackage{esvect}
\usepackage[usenames, dvipsnames]{color}
\usepackage{epsfig}
\usepackage{graphicx}
 \usepackage{float} 
\usepackage{gensymb}
\usepackage{multirow}
\usepackage{dcolumn}   
\usepackage{bm}        
\usepackage{ulem}
\begin{document}

\title{Emergence of well screened states in a superconducting material of the CaFe$_2$As$_2$ family}

\author{Ram Prakash Pandeya,$^1$ Anup Pradhan Sakhya,$^1$ Sawani Datta,$^1$ Tanusree Saha,$^2$ Giovanni De Ninno,$^{2,3}$ Rajib Mondal,$^1$ C. Schlueter,$^4$ A. Gloskovskii,$^4$ Paolo Moras,$^5$ Matteo Jugovac,$^5$ Carlo Carbone,$^5$ A. Thamizhavel$^1$ and Kalobaran Maiti$^1$}
\altaffiliation{Corresponding author: kbmaiti@tifr.res.in}

\affiliation{$^1$Department of Condensed Matter Physics and Materials Science, Tata Institute of Fundamental Research, Homi Bhabha Road, Colaba, Mumbai - 400005, India}
\affiliation{$^2$Laboratory of Quantum Optics, University of Nova Gorica, 5001 Nova Gorica, Slovenia}
\affiliation{$^3$Elettra-Sincrotrone Trieste, Area Science Park, 34149 Trieste, Italy}
\affiliation{$^4$Deutsches Elektronen-Synchrotron DESY, 22607 Hamburg, Germany}
\affiliation{$^5$Istituto di Struttura della Materia, Consiglio Nazionale delle Ricerche, Area Science Park, I-34149 Trieste, Italy}

\date{\today}

\begin{abstract}
Coupling among conduction electrons (e.g. Zhang-Rice singlet) are often manifested in the core level spectra of exotic materials such as cuprate superconductors, manganites, etc. These states are believed to play key roles in the ground state properties and appear as low binding energy features. To explore such possibilities in the Fe-based systems, we study the core level spectra of a superconductor, CaFe$_{1.9}$Co$_{0.1}$As$_2$ (CaCo122) in the CaFe$_2$As$_2$ (Ca122) family employing high-resolution hard $x$-ray photoemission spectroscopy. While As core levels show almost no change with doping and cooling, Ca 2$p$ peak of CaCo122 show reduced surface contribution relative to  Ca122 and a gradual shift of the peak position towards lower binding energies with cooling. In addition, we discover emergence of a feature at lower binding energy side of the well screened Fe 2$p$ signal in CaCo122. The intensity of this feature grows with cooling and indicate additional channels to screen the core holes. The evolution of this feature in the superconducting composition and it's absence in the parent compound suggests relevance of the underlying interactions in the ground state properties of this class of materials. These results reveal a new dimension in the studies of Fe-based superconductors and the importance of such states in the unconventional superconductivity in general.
\end{abstract}

\maketitle

\section{Introduction}

Origin of high temperature ($T_c$ = transition temperature) superconductivity continues to be an outstanding issue in material science. These materials are called unconventional superconductors due to the anisotropy in superconducting gap, effective two-dimensionality in the charge carrier conduction, pseudogap phase and a deviation from the Fermi liquid behavior above $T_c$, etc. Various spectroscopic and theoretical studies of the cuprate superconductors \cite{CuScs-Damascelli,CuScs-Kastner} show signature of singlet states such as Zhang-Rice singlet \cite{ZR} in their electronic structure. These states are believed to be important for the formation of Cooper pairs responsible for superconductivity. Discovery of Fe-based superconductors \cite{Kamihara,Hosono,Stewart} added additional complexity due to the observation of several unusual phenomena; e.g. coexistence of magnetism and superconductivity, nematicity, pressure induced rich phase diagram, etc. In contrast to an effective single band problem in cuprates, multiple bands cross the Fermi level, $\epsilon_F$ in Fe-based systems that makes the detection of various exotic features difficult.

We studied the electronic structure of a superconducting Fe-based system, CaCo122 employing bulk sensitive, hard $x$-ray photoemission spectroscopy (HAXPES). The results are compared with the high-resolution ultraviolet photoemission spectroscopy (UPS) data to identify the surface contributions. The experimental results of CaCo122 are compared with those of the parent compound, Ca122 which does not show superconductivity at the atmospheric pressure. Ca122 forms in the tetragonal structure at room temperature and undergoes a magneto-structural transition at 170 K to orthorhombic spin-density wave state.\cite{Ni,Kreyssig,Goldman} The properties of Ca122 can be varied smoothly by application of pressure and/or substitution of Fe with other transition metal elements such as Ni, Co, Rh  \textit{etc}.\cite{Crystal,Ca122-Harnagea,Ca122-Danura,Qi} The Co-doped compound, CaCo122 remains in the tetragonal phase in the whole temperature range studied, does not show magnetic transition and superconductivity appears below 15 K. We discover a new feature at the lower binding energy side of the Fe 2$p$ HAXPES data of the superconducting composition which is absent in the parent compound indicating it's relevance in the ground state properties.

\section{Experimental Technique}

High quality single crystals of Ca122 and CaCo122 were grown using flux method.\cite{Crystal} Laue diffraction pattern of both the materials show sharp spots indicating high degree of single crystallinity. Samples were characterized using $x$-ray diffraction (XRD) and energy dispersive analysis of $x$-rays, and were found to be in single phase with no impurity feature. Structural transition of Ca122 is well documented in the literature.\cite{Ni,Kreyssig,Goldman} Temperature dependent XRD measurements of CaCo122 show tetragonal structure in the whole temperature range of 4 - 300 K studied. Lattice parameters show small gradual decrease due to thermal compression. Magnetic susceptibility measurements show transition to diamagnetic behavior below 15 K suggesting onset of superconductivity; these data are given in the supplemental materials.\cite{suppl}

HAXPES measurements were carried out using 6 keV photon energy at P-22 beamline, Petra-III, DESY, Hamburg \cite{HAXPES_Book_Ref1} and UPS was done at the VUV beamline, Elettra, Trieste. The samples were cleaved in ultrahigh vacuum using top-post removal method. The overall energy resolution for HAXPES and UPS measurements were 200 meV and 10 meV, respectively. To consider polarization induced effects in the valence band spectra,\cite{takegami_PRB19} the measurements were carried out in the normal emission geometry. UPS studies are done only for As 3$d$ emissions and the incident light comes at an incidence angle of 45$^\circ$ with the sample surface normal. For HAXPES measurements (incidence angle is about 85$^\circ$), the light polarization was aligned close to the detector axis (out-of-plane polarization is about 99.7\%) ensuring large contributions from $d_{xz}$/$d_{yz}$ orbitals which are strongly hybridized with the As 4$p$ states and have major contributions in the ground state properties.\cite{Ca122-Ram}

\section{Results and Discussions}

\begin{figure}
\includegraphics[width=0.95\linewidth]{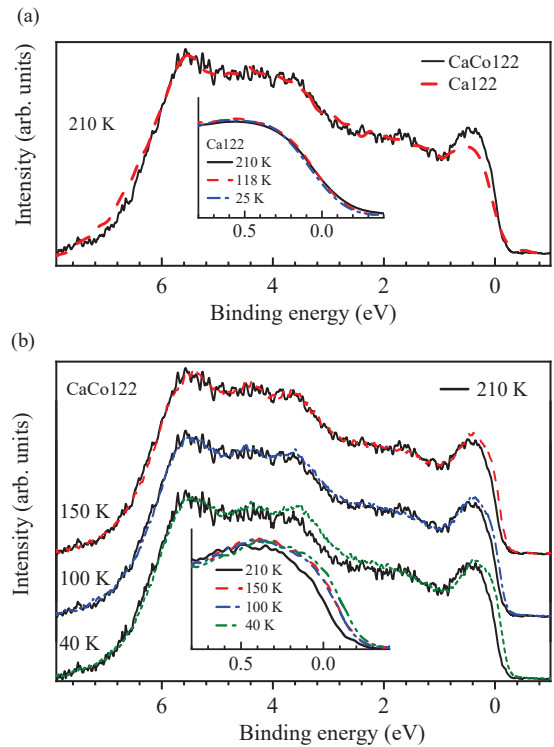}
\caption{(Color online) (a) Valence band spectra of CaCo122 (solid line) and Ca122 (dashed line) at 210 K. Inset shows temperature dependence of the near $\epsilon_F$ peak in Ca122. (b) Valence band spectra of CaCo122 at different temperatures. 210 K data (solid line) is superimposed on the data in every case. Inset shows the temperature evolution of the near $\epsilon_F$ feature of CaCo122.}
\label{FIG1VBW}
\end{figure}

In Fig. \ref{FIG1VBW}, we show the valence band spectra of CaCo122 (solid line) and Ca122 (dashed line) collected at 210 K after subtracting Shirley background in each case. Both the spectra show several distinct features consistent with the reported data for the parent compound.\cite{Ganesh-Ca122,Ca122-Ram} The features between 3-4 eV binding energies are primarily contributed by the As 4$p$ states. The higher binding energy features are the bonding states with large As 4$p$ contributions. The antibonding states appear near $\epsilon_F$ having dominant Fe 3$d$ character.\cite{Ganesh-Ca122} Interestingly, while the intensities of the features beyond 1 eV binding energy are almost identical in both the cases, the intensity of the feature near $\epsilon_F$ is significantly enhanced in CaCo122. There are several reasons that can enhance intensity. Co-substitution dopes electrons into the system. The photoemission cross section of Co 3$d$ states is larger than Fe 3$d$ cross section; the atomic photoemission cross-section of Co 3$d$ states is about 1.67 times of the Fe 3$d$ cross-section at 6 keV.\cite{cross-section} This may have an effect as the valence states are expected to have Co 3$d$ contributions. There could be a change in the orbital character of the bands close to $\epsilon_F$ due to a change in covalency via Co-substitution. Clearly, Co-substitution essentially enhances the intensity below $\epsilon_F$ without observable shift of $\epsilon_F$.

Decrease in temperature has significant effect in the electronic structure of the superconducting CaCo122 while the HAXPES data of the parent composition does not show detectable change [see inset in Fig. \ref{FIG1VBW}(a)].\cite{Ca122-Ram} The spectral change in CaCo122 is shown in the inset of Fig. \ref{FIG1VBW}(b) in an expanded energy scale. Clearly, an additional feature near $\epsilon_F$ grows in intensity with cooling. The magneto-structural transition in Ca122 leads to a change in the local structural parameters.\cite{EXAFS-Ram} CaCo122 having no magneto-structural transition also show similar behavior. However, a link between the spectral changes and structural transition can be ruled out based on the following observations: ARPES studies show opening of a gap at $\epsilon_F$ due to the formation of spin-density wave (SDW) state in Ca122 below 170 K \cite{Khadiza-ARPES} which shows a marginal reduction of the spectral weight in Ca122 shown in the inset of Fig. \ref{FIG1VBW}(a). This is opposite to the scenario observed in CaCo122. Evidently, the temperature induced spectral evolution in CaCo122 on cooling is interesting and may have an implication in the ground state properties of this system.\cite{yamasaki}

\begin{figure}
\includegraphics[width=0.95\linewidth]{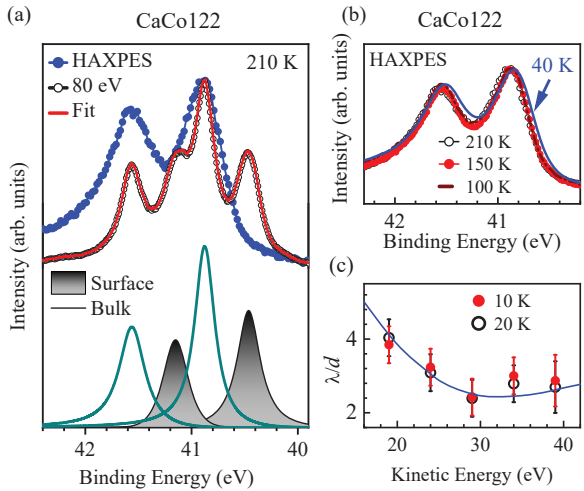}
\vspace{-24ex}
\caption{(a) As 3\textit{d} core level spectra at 210 K measured using 80 eV and 6 keV photon energies. The simulated UPS data is shown by line superimposed on the experimental data. The component peaks are shown in the lower panel. (b) As 3\textit{d} HAXPES data at different temperatures. (c) Estimated $\lambda/d$ values as a function of photoelectron kinetic energy. The smooth curve drawn over the data points is a guide to eye.}
\label{Fig2As3d}
\end{figure}

In order to investigate the surface effect in the electronic structure, we chose As 3$d$ core levels which can be probed using high resolution UPS technique in addition to HAXPES. As 3$d$ spectra of CaCo122 collected using 6 keV and 80 eV photon energies are shown in Fig. \ref{Fig2As3d}(a). HAXPES data exhibit two intense features at 40.8 eV and 41.5 eV binding energies for the spin-orbit split 3$d_{5/2}$ and 3$d_{3/2}$ signals. The 80 eV data exhibit two additional intense features at the lower binding energy sides (40.5 eV and 41.2 eV). At 80 eV, the kinetic energy of As 3$d$ photoelectrons will be about 34 eV which is close to the lowest escape depth. Thus, the additional features in the 80 eV data are attributed to the surface electronic structure as also observed in other Fe-based 122-compounds.\cite{Jong,Ganesh-Ca122} In Fig. \ref{Fig2As3d}(b), the HAXPES data at different temperatures are shown exhibiting identical lineshape and spectral features down to 100 K. The 40 K data is weakly shifted towards lower binding energies along with a small enhancement of the asymmetry.

After identifying the surface and bulk features, we extract the information about the photoelectron escape depth, $\lambda$ and the surface layer thickness, $d$. We fit the UPS spectra ($h\nu$ = 60, 65, 70, 75 and 80 eV) using a set of asymmetric Gaussian-Lorentzian (GL) product functions; a representative fit of the 80 eV spectrum is shown in the Fig. \ref{Fig2As3d}(a). The binding energies and the full width at half maximum (FWHM) of the core level peaks are given in the supplemental material.\cite{suppl} Simulated data provide an excellent representation of the experimental spectrum with the intensity ratios of the spin-orbit split features equal to the ratio of their multiplicities. The energy difference between the surface and bulk peaks is found to be about 400 meV and the spin-orbit splitting is found to be 700 meV as also evident in the raw data.

The photoemission intensity, $I(\epsilon)$ can be expressed as $I(\epsilon)=\int_{0}^{d}I^{s}(\epsilon)e^{-x/\lambda}dx + \int_{d}^{\infty} I^{b}(\epsilon)e^{-x/\lambda}dx$
where, $I^{s}(\epsilon)$ and $I^{b}(\epsilon)$ are the surface and bulk contributions to the total intensity. Thus, the intensity ratio of the surface and bulk peaks extracted from the core level analysis, is equal to $(e^{d/\lambda} -1)$. This relation is used to extract the $\lambda/d$ values.\cite{Arindam,Pincelli_lambda_by_d} The results obtained from the 10 K and 20 K data are shown in Fig. \ref{Fig2As3d}(c) exhibiting a minima between 30 - 40 eV similar to the universal curve.\cite{UniversalCurve} Usually, $\lambda$ is about 20 \AA\ in Al $K\alpha$ energies and becomes 6 - 8 \AA\ at the highest surface sensitivity.\cite{surf} Thus, a typical value of $d$ appears to be 3 - 4 \AA\ in this material. Since the sample is known to cleave at the Ca-layer leaving about 50\% of Ca atoms on both the cleaved surfaces, the surface contribution appears to be limited within the top Ca-As-Fe layers. The HAXPES data represents essentially the bulk electronic structure.

\begin{figure}
\vspace{4ex}
\includegraphics[width=0.95\linewidth]{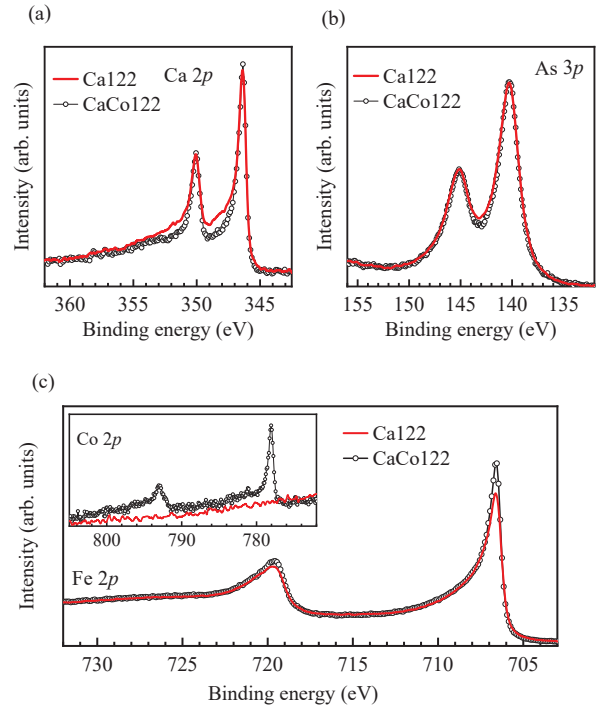}
\vspace{-4ex}
\caption{HAXPES spectra of (a) Ca 2$p$, (b) As 3$p$ and (c) Fe 2$p$ core level excitations at 210 K. The inset in (c) is the Co 2$p$ spectrum. The symbols represent the CaCo122 data and the lines are the Ca122 data.}
\label{Fig3Core}
\end{figure}

The doping induced changes in the core level spectra is investigated in Fig. \ref{Fig3Core}. Ca 2$p$ HAXPES spectra are shown in Fig. \ref{Fig3Core}(a). The intense sharp peaks at 346.5 eV and 350.5 eV are the bulk peaks, and the shoulders at 348 eV and 352 eV are the surface features for 2$p_{3/2}$ and 2$p_{1/2}$ photoemissions, respectively.\cite{Ca122-Ram} We observe a depletion of intensity at about 348 and 352 eV due to doping which indicates that surface-bulk differences in the Ca contributions of the electronic structure is reduced in the doped composition. This is significant as the Ca states have significant hybridization with the valence states which is expected to play a role (charge reservoir layer) in the superconductivity of this material.\cite{Khadiza-DFT} As 3$p$ spectra shown in Fig. \ref{Fig3Core}(b) exhibit spin-orbit split sharp and distinct features at 140 eV and 145 eV with a splitting of about 5 eV. Co substitution does not have any visible effect in the spectra.

In Fig. \ref{Fig3Core}(c), we study the Fe 2$p$ spectra of Ca122 and CaCo122 samples; Co 2$p_{3/2}$ and 2$p_{1/2}$ signals appear at 778 and 793 eV [see the inset of Fig. \ref{Fig3Core}(c)] and confirms Co-doping. Fe 2$p$ well screened features appear at 706.5 eV and 719.5 eV binding energies for 2$p_{3/2}$ and 2$p_{1/2}$ excitations. In metals and in Fe-based systems as well, the poorly screened features at higher binding energies are weak and difficult to distinguish from the background intensities. A normalization of the spectra by the intensities at the higher binding energy region show significant enhancement of the relative intensity of the well-screened features in Co-doped sample. This suggests that the core hole is more efficiently screened in the superconducting composition in comparison to the parent compound.

\begin{figure}
\vspace{4ex}
  \includegraphics[width=0.95\linewidth]{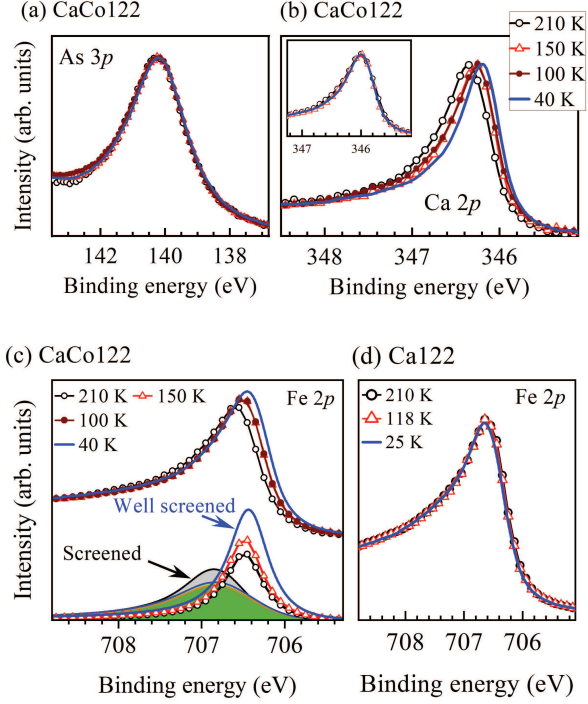}
\vspace{-4ex}
  \caption{HAXPES spectra of (a) As 3\textit{p}, (b) Ca 2\textit{p}, and (c) Fe 2\textit{p} core levels measured using 6 keV photon energy at different temperatures. Inset in (b) is obtained after shifting the 150 K, 100 K and 40 K spectra by 80, 80 and 140 meV towards high binding energies, respectively. Spectra shown in the lower panel of (c) are the simulated Fe 2\textit{p} signal. (d) Fe 2$p$ spectra of Ca122 at different temperatures.}
  \label{Fig4Temp}
\end{figure}

In order to probe the effect of temperature on the electronic structure, we study the spectra collected at different temperatures in Fig. \ref{Fig4Temp}. The As 3\textit{p} core level spectra are almost identical at all the temperatures. Clearly, the phonon contributions and/or recoil effect are not significant relative to the intrinsic width of the peaks. Ca 2$p_{3/2}$ spectra show a gradual shift to lower binding energies with cooling; the 150 K and 100 K data superimposes on each other suggesting negligible thermal effect in this temperature range. We have shifted the 150 K spectrum by 80 meV and 40 K spectrum by 140 meV, and show them in the inset of Fig. \ref{Fig4Temp}(b). All the features superimposes almost perfectly. Such a rigid energy shift towards lower binding energies with cooling suggest effective reduction of the Madelung potential at the Ca sites without significant change in it's properties. A significant decrease in As-Ca bond lengths is observed with decreasing temperature\cite{EXAFS-Ram} suggesting enhanced hybridizations in this temperature range which is in line with the core level shift observed here. It is to note here that although the spectral evolution with temperature down to 100 K looks in line with the change in lattice constant, the understanding of the large change observed in the 40 K data needs further study.

The Fe 2$p_{3/2}$ spectra of CaCo122 shown in Fig. \ref{Fig4Temp}(c) exhibit puzzling changes while the Fe 2$p_{3/2}$ signal of Ca122 shown in Fig. \ref{Fig4Temp}(d) are identical at all the temperatures. The growing intensities at the lower binding energy side of the screened peak suggests emergence of a new feature which becomes stronger with cooling. In order to distinguish these two features, we have named them `{\it screened peak}' and `{\it well screened peak}'. In order to get a qualitative idea on underlying physics, let us note that in cuprate superconductors, the screened Cu 2$p$ HAXPES spectra exhibit multiple features depending on the behavior of the holes in the valence band. One peak correspond to core hole screening via hopping of an electron from the ligand levels to the photoemission site which is the `screened peak'. There is a second peak at lower binding energies where the hole in the ligand band forms a singlet with the neighboring Cu 3$d$ hole, termed as Zhang-Rice singlet\cite{ZR,Taguchi_LCO-NCO,BoeskePRB} providing additional stability of the corresponding final state. The HAXPES data from manganites also show signature of additional feature at the lower binding energy side of the screened peak due to the final state where screening is done by conduction electrons.\cite{Horiba_LSMO,TVR} A similar scenario may be operative in the present case too.

To understand the final state effect in the photoemission spectra of varied correlated electron systems, the typical multiband Hubbard Hamiltonian \cite{Ni-Dimen} used is as follows:
$H=\epsilon_{d\alpha}d_{\alpha\sigma}^\dagger d_{\alpha\sigma}+\epsilon_pp_{i\sigma}^\dagger p_{i\sigma}+\epsilon_c{c_{\mu\sigma}^\dagger}c_{\mu\sigma}+[t_{dp}^{\alpha i}d_{\alpha\sigma}^\dagger p_{i\sigma}+h.c.]+[t_{pp}^{ij }p_{i\sigma}^\dagger{p_{j \sigma}+h.c.}]
+U_{dd}^{\alpha\beta\gamma\delta}d_{\alpha\sigma_1}^\dagger d_{\beta\sigma_2}^\dagger{d_{\gamma\sigma_3}}d_{\delta\sigma_4}
+U_{dc}^{\alpha\beta\mu\nu}d_{\alpha\sigma_1}^\dagger d_{\beta\sigma_2}c_{\mu\sigma_3}^\dagger c_{\nu\sigma_4}
+\zeta_{2p}\langle\mu|l.s|\nu\rangle{c_{\mu\sigma}^\dagger}{c_{\nu\sigma'}}$
%
where, repeated indices indicate sum and the Fermion operators $d_{\alpha\sigma}^\dagger$, $p_{i\sigma}^\dagger$ and $c_{\mu\sigma}^\dagger$ represent the creation of an electron in Fe 3$d$, As 4$p$ and Fe 2$p$ levels respectively. $U_{dd}^{\alpha\beta\gamma\delta}$ and $U_{dc}^{\alpha\beta\mu\nu}$ are the Coulomb repulsion strengths among Fe 3$d$ electrons, and between Fe 2$p$ and Fe 3$d$ electrons, respectively. For a simplistic description, the eigenstates can be described within a configuration interaction picture considering a linear combination of unscreened $|d^n\rangle$ ($n$ is the number of 3$d$ electrons) and screened $|d^{n+1}\underbar{p}\rangle$ states where $|\underbar{p}\rangle$ is a As 4$p$ hole state. If $|\underbar{p}\rangle$ forms a singlet with a 3$d$ hole, $|\underbar{d}\rangle$ at the neighboring Fe site, the energy of the corresponding state, $|S\rangle = (|\underbar{p}_\uparrow\underbar{d}_\downarrow\rangle - |\underbar{p}_\downarrow\underbar{d}_\uparrow\rangle)$ will be lowered by the binding energy of the singlet state. If this state is called $|\underbar{p}^\star\rangle$, the corresponding final state would be $|d^{n+1}\underbar{p}^\star\rangle$. Thus, the problem can be reduced to an effective three states problem with the basis $|d^n\rangle$, $|d^{n+1}\underbar{p}\rangle$ and $|d^{n+1}\underbar{p}^\star\rangle$. In cuprates, the energy difference between the screened peaks with the hole in the ligand band without singlet formation and the hole formed a Zhang-Rice singlet is about 1.5 eV.\cite{Taguchi_LCO-NCO} The energy separation of the features in manganites due to two types of screening is also close to 1.5 eV.\cite{Horiba_LSMO} Here, we observe an energy separation of about 0.4 eV. This is in line with the expected behavior. In both cuprates and manganites, the hybridization of the transition metal $d$ and oxygen $p$ states is strong due to the planer geometry. However, in this Fe-based system, the Fe layers are sandwiched by two As layers; they do not lie in the same plane. Fe-As-Fe angle is about 71$^\circ$ - 74$^\circ$ in these systems.\cite{Santa_BondAngle} Thus, the hybridization of the As 4$p$ holes created due to the core-hole screening at the Fe-sites will have significantly weaker hybridization with the neighboring Fe-sites compared to the cases in cuprates and manganites.\cite{hopping_PRB09} With the decrease in temperature, the Fe-As bondlength reduces in addition to reduction of thermal vibrations. Therefore, these states will be more stable at lower temperatures as observed in various other systems. Signature of such an enhancement of $|\underbar{p}^\star\rangle$ states with cooling is consistent with the enhancement of intensity observed in the valence band spectra. It is to note here that all these observations are carried out above the superconducting transition temperature. The presence of these features in the superconducting compositions and their absence in the parent compound indicate their relevance in the ground states properties like cuprates although these features may occur due to the presence of highly itinerant states as found in manganites. More studies are necessary to understand the link between the superconductivity and the behavior of these states.

\section{Conclusion}

In conclusion, we have studied the final state effects in the core level spectroscopy of the bulk electronic structure of CaFe$_2$As$_2$ and its Co-doped superconducting composition, CaFe$_{1.9}$Co$_{0.1}$As$_2$ employing hard $x$-ray photoemission spectroscopy. Core level spectra related to As do not show visible change with temperature. Ca 2$p$ spectra exhibit reduced surface-bulk differences in the doped composition and a shift towards lower binding energies with the decrease in temperature. In addition, we discover a new feature in the Fe 2$p$ core level spectra at the lower binding energy side of the screened peak. Intensity of this feature increases with the decrease in temperature. This is attributed to the final state where the As 4$p$ hole created due to the hopping of an As 4$p$ electron to Fe 3$d$ at the photoemission site to screen the core hole, forms a singlet with the transition metal 3$d$ hole as found in cuprates and manganites. The emergence of such a feature in the superconducting composition and its growth at low temperatures indicate a possible link to the superconductivity in this class of materials and calls for further studies in this direction.

\section{Acknowledgement}

Authors acknowledge the financial support under India-DESY program and Department of Atomic Energy (DAE), Govt. of India (Project Identification no. RTI4003, DAE OM no. 1303/2/2019/R\&D-II/DAE/2079 dated 11.02.2020). K.M. acknowledges financial assistance from DAE under the DAE-SRC-OI Award program.

\end{document}